# Advances in neutral tungsten ultraviolet spectroscopy for the potential benefit to gross erosion diagnosis

**C. A. Johnson[1], D. A. Ennis[1], S. D. Loch[1], G. J. Hartwell[1], D. A. Maurer[1], S. L. Allen[2], B. S. Victor[2], C. M. Samuell[2], T. Abrams[3], E. A. Unterberg[4], R. T. Smyth[5]**

[1]Auburn University, Auburn, AL, 36849, United States of America
[2]Lawrence Livermore National Laboratory, Livermore, CA 94550, United States of America
[3]General Atomics, San Diego, CA, 92186, United States of America
[4]Oak Ridge National Laboratory, Oak Ridge, TN 37831, United States of America
[5]Queens University Belfast, Belfast BT7 1NN, Northern Ireland, UK

E-mail:

13 February 2019

**Abstract.**
    A spectral survey of tungsten emission in the ultraviolet region has been completed in the DIII-D tokamak and the CTH torsatron to assess the potential benefit of UV emission for the diagnosis of gross W erosion. A total of 29 W I spectral lines are observed from the two experiments using survey spectrometers between 200-400 nm with level identifications provided based on a structure calculation for many of the excited states that produce strong emission lines. Of the 29 observed lines, 20 have not previously been reported in fusion relevant plasmas, including an intense line at 265.65 nm which could be important for benchmarking the frequently exploited line at 400.88 nm.   Nearly all of the observed spectral lines decay down to one of the six lowest energy levels for neutral W, which are likely to be long-lived metastable states. The impact of metastable level populations on the W I emission spectrum and any erosion measurement utilizing a spectroscopic technique is potentially significant. Nevertheless, the high density of W I emission in the UV region allows for the possibly of determining the relative metastable fractions and plasma parameters local to the erosion region. Additionally, the lines observed in this work could be used to perform multiple independent gross erosion measurements, leading to more accurate diagnosis of gross tungsten erosion.

## 1. Introduction

Tungsten is a leading candidate for divertor Plasma Facing Components (PFC) due to its excellent thermal properties, low sputtering yield and low hydrogen solubility [1]. The erosion of tungsten PFC will play a critical role in determining the plasma stability and performance of ITER where core concentrations of W as low as $10^{-5}$ could result in significant radiative losses and degrade plasma performance [2, 3]. Furthermore, wall erosion could result in up to thousands of kilograms per year of circulating material in a power reactor [4]. Therefore, accurate knowledge of PFC total erosion rates is essential in present and future experiments to help identify dominant sources of impurities from wall components and understand core impurity concentrations. A number of fusion devices have investigated tungsten as a PFC material including: the ASDEX-Upgrade which converted to a solid W divertor [5], the JET tokamak having an ITER-like W monoblock divertor [6] and TEXTOR which previously conducted experiments with W limiters [7–9]. Recently, experiments were conducted in the DIII-D tokamak with tungsten coated titanium-zirconium-molybdenum inserts in the divertor to investigate tungsten sourcing and transport [10, 11].

Gross erosion of PFC material can be diagnosed from neutral spectral line emission together with S/XB atomic coefficients representing the 'ionizations per photon' [12–14]. Gross erosion represents the total erosion that occurs; however, many of the eroded atoms promptly redeposit back onto the material surface and do not remain in the plasma. The net erosion characterizes the number of eroded atoms that do not promptly redeposit and is a small fraction of the gross erosion for tungsten [15,16]. The net erosion can potentially be determined from emission of multiple charge states in combination with the relevant S/XB coefficients.

For light atomic systems ($Z < 10$) it is well known that the metastable level fraction is important to accurately predict the spectral emission [17]. Metastable resolved transport coefficients (such as effective ionization and recombination rate coefficients) are needed to accuracy predict the transport of light system impurities [18]. It is likely that metastable fractions are also important in the modeling of erosion rates and transport for heavy atomic systems. Yet, the impact of metastable populations on the W I emission spectrum has not been well documented, to date, as part of erosion measurements utilizing the S/XB technique. Previous theoretical modeling by Beigman et al. [19], varied the metastable populations of neutral tungsten using a thermodynamic temperature $T_W$ that fixed the metastable populations to local thermodynamic equilibrium values and showed that the S/XB ratio for the 400.88 nm W I line varied by up to a factor of five for the $T_W$ values that were varied. The only way to account for metastable level contributions is to simultaneously measure numerous W I spectral lines driven from different metastable states. The ideal scenario to determine the W I metastable fractions would utilize ratios of lines with the same temperature and density behavior but different dependencies on the metastable driving population. In principle, spectral lines could be exploited to determine ratios of the ground and all metastable populations that drive emission in each of these lines which in turn could yield more accurate erosion measurements. However, this metastable analysis requires an accurate atomic model to account for temperature and density dependencies of each transition. Improved calculations of W I atomic data could assist with such investigations [20]. One principal aim of this paper is to specify a set of W

I spectral lines that could be utilized to diagnosis the relative metastable fraction and provide information about the metastable contribution. Determining the dependence of W I emission on the metastable populations is an important step toward future non-invasive, real-time spectral diagnosis of gross and net erosion.

Ultraviolet survey spectrometers (200 to 400 nm) have been employed on both the DIII-D [21] and Compact Toroidal Hybrid (CTH) [22] machines during experiments with tungsten sources for the purpose of enhancing the diagnosis of gross tungsten erosion in edge plasmas. Neutral tungsten emission is predicted to be most intense in the ultraviolet range [23] making UV spectroscopy attractive for total erosion measurements. Simultaneously acquiring multiple independent gross erosion measurements is possible where multiple W I emission lines are clustered within a narrow wavelength range [24]. The impact of metastable states on the W I emission spectrum and S/XB ratios could be significant and observing multiple lines should allow the metastable fractions to be determined. Furthermore, intensity ratios of appropriately selected W spectral lines can provide separate measures of electron temperature and density in the region where erosion is occurring.

The setup and operation of new ultraviolet survey spectrometers installed on both the DIII-D tokamak and CTH torsatron are detailed in Section 2. Ultraviolet spectra from DIII-D and CTH are presented including the first measurements of several intense W I lines in fusion relevant plasma conditions. Complete identification of the observed W I spectral transitions is presented in Section 3. Finally, recommendations for clusters of W I spectral lines and wavelength ranges in the UV to account for the effects of metastable level populations yielding more accurate gross and net erosion measurements and improved tests of W I erosion models are discussed in Section 4.

## 2. Experimental Method

In the DIII-D tokamak, tungsten was introduced into the divertor region by installing two continuous rings of tungsten coated inserts on the divertor shelf and floor [25]. Tungsten emission is observed with passive UV spectroscopy with the collection optics for the survey spectrometers located at a toroidal angle of 75° and poloidally near the top of the machine (75R+2). Two UV transmitting, 78 mm lenses provide simultaneous measurements in the divertor region that can be independently aligned to either the divertor shelf or divertor floor. The optical lines of sight for the spectrometers are shown in Figure 1 overlaid on an equilibrium reconstruction of magnetic flux surfaces for DIII-D shot 167350 at 1935 ms. The location of a single Langmuir probe used to determine the electron temperature and density at the tungsten surface is shown in the inset. The line-of-sight terminating at the floor ring is partially obstructed by the divertor nose due to the location of the collection optics. As seen in the insert of Figure 1, the collection spot size on the divertor tiles is 3.3 cm in diameter for both the divertor floor and shelf, significantly larger than the W I ionization scale length ($\sim$ mm) and totally encompassed within the 5 cm wide W coated tile insert. UV transmitting lenses and 1000 $\mu$m diameter, 5 m long fused-silica fibers were selected to maximize transmitted light as UV wavelengths are severely attenuated along optical fibers. Two StellarNet EPP2000 spectrometers [26] sensitive between 200-300 nm and 300-400 nm were used on both DIII-D and CTH for the work presented here. Each spectrometer has a crossed Czerny-Turner design with a 2400 grooves/mm ruled plane grating, an entrance slit of 7 $\mu$m, and a resolution of $\sim$ 0.1 nm. Both spectrometers utilize a Sony ILX511 CCD detector with 2048 pixels. During DIII-D experiments the UV survey spectrometers were housed in a neutron and x-ray shield box. Spectrometer exposure times range from 30 to 2000 ms but typical W signal levels in DIII-D require integration times greater than 100 ms. A wavelength calibration over the entire range of the spectrometers is accomplished using Hg-Ne, Hg-Ar, and Zn pen lamps. The resolving power of the UV survey spectrometers make it possible to identify W I spectral lines that have potential diagnostic utility. Absolute calibration of the spectrometer sensitivity is necessary to measure W erosion via the S/XB method or to compare line intensities as the system's photon efficiency can vary substantially with wavelength across the UV range. However, an in-situ absolute intensity calibration of the UV survey spectrometers was limited by available calibration sources to 350-400 nm.

Tungsten is introduced into the CTH experiment locally using a vertically translating W-tipped probe. The probe is constructed with a 0.75" diameter 1.0" long cylindrical tungsten tip backed by a 9.5" long boron nitride sleeve. The CTH device is a five-fold symmetric torsatron [22] limited at five locations by C, Mo and stainless steel. The tungsten probe can be fully retracted from the plasma or inserted to varying depths. When the probe is inserted to 6 cm, the W probe tip is well within the last closed flux surface of the plasma, corresponding to a normalized toroidal flux of approximately 0.6 (Figure 2). Additionally, the vertical position of the plasma can be adjusted to move the plasma away from the probe tip by use of a radial field coil. When the probe is inserted into the plasma, all emission lines increase, which is attributed to additional plasma-probe interaction. The increase of background impurity lines not resulting from erosion of the probe makes correctly identifying lines more challenging. The UV survey spectrometer collection optics are located directly opposite from the W probe as shown in Figure 2. A UV lens of focal length 50 mm is used in combination with a 1.6 m fused silica fiber to produce a collection spot size equal to that of the probe tip cross section (0.75" diameter). The shorter distance between the collection optics and the W source and consequently larger solid angle (as compared to the DIII-D optical setup) results in more intense and less polluted measurements of W spectral lines.

The limited spectral resolution of the UV survey spectrometers ($\sim$ 0.1 nm) meant that spectral lines from other impurities could be blended with W lines of interest. Thus, methods in addition to wavelength agreement with the NIST Atomic Spectra Database [27] were needed to positively identify lines resulting from W in contrast to background impurity lines. One technique utilized in the CTH experiment to identify W spectral lines involved probe depth scans. As the W probe tip is inserted, plasma-probe interactions should increase as more W is eroded; thus, line intensities from W

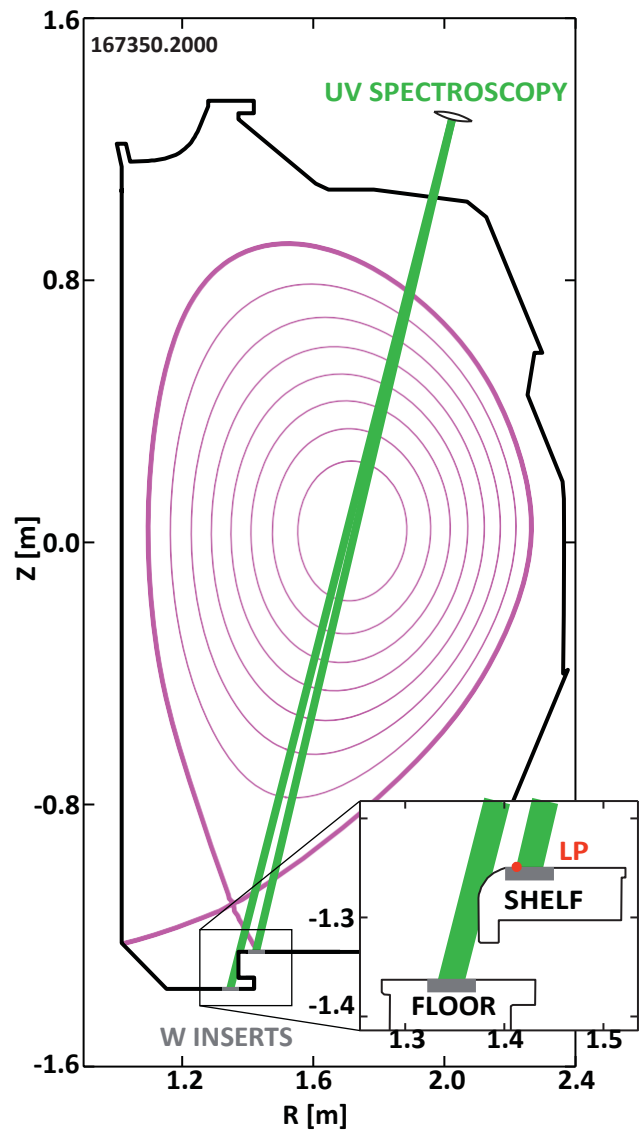

**Figure 1.** UV survey spectrometer lines of sight viewing the DIII-D tokamak divertor highlighted in green and overlaid on a reconstruction of magnetic flux surfaces. Two toroidally displaced UV lenses provide independent views that can be aligned to tungsten coated inserts on either the divertor floor or shelf. The location of a Langmuir probe on the divertor shelf is shown in red. Figure adapted from Abrams et al. [11]

should intensify more than other intrinsic impurities in CTH plasmas. Some background impurity line intensities also increase with probe insertion depth but not at the same rate as W lines. Additionally, stainless steel and Mo-tipped probes also yield spectra for comparison to determine base impurity emission increases with probe insertion. The probe experiments in CTH allow for the W source to be completely removed from the plasma. Lines that increase in intensity when both the stainless steel and Mo probes are inserted are assumed to be CTH intrinsic impurities or lines coming from the boron nitride probe sleeve. For measurements in DIII-D, ratios of lines are examined to distinguish intrinsic impurities lines from W spectral lines. As the strike point sweeps

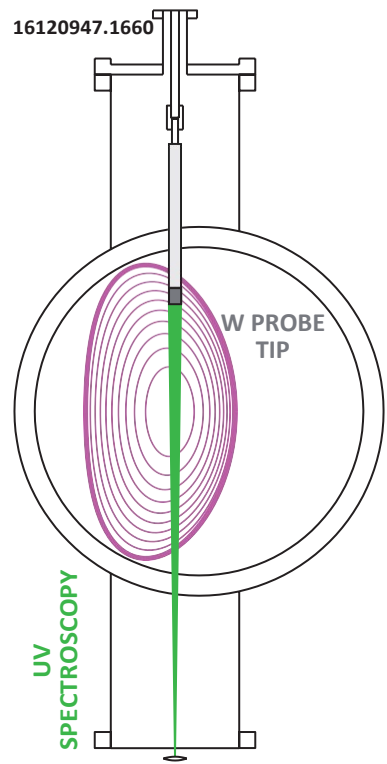

16120947.1660

**Figure 2.** UV survey spectrometer line of sight viewing the tungsten-tipped probe highlighted in green and overlaid on a reconstruction of magnetic flux surfaces. The W probe is located directly above the spectrometer collection optics and the W tip is well within the last closed flux surface when fully inserted.

from carbon tiles to W inserts, the ratio of spectral intensities from W to C should increase. The ratio of spectral lines from background impurities in DIII-D compared to C lines does not increase when the strike point sweeps over a W insert.

The line identification procedure assigns each observed tungsten spectral line with a rating of either A, B or C to quantify the confidence in the line identification. Lines observed in only one of the experiments are given a 'C' rating. Lines observed in only one of the experiments and identified in the NIST Atomic Spectra Database as being intense or lines observed in both machines were given a 'B' rating. Lines observed in both experiments and also identified in the NIST Atomic Spectra Database as being intense were given an 'A' rating. However, spectral lines with an 'A' rating are still possibly blended with other base impurities which do not appear in the CTH spectrum when either the Mo or stainless steel probes are inserted. Therefore, while lines with an 'A' rating indicate the highest identification confidence, it remains possible any of the lines are contaminated to some degree by impurity emission.

## 3. Experimental Results

Tungsten emission lines are among the least intense impurity UV lines present in measured DIII-D spectra, consistent with the low sputtering yield of tungsten [28]; yet,

W I spectral line identification could still be accomplished. Other intrinsic impurities such as carbon and boron dominate the UV spectrum. As mentioned, the limited resolution of the UV survey spectrometers may lead to overlap of other impurity emission lines with the W lines. The possibility of overlapping emission necessitated the use of the metrics explained in Section 2 to positively identify W spectral lines.

Tungsten emission is more intense compared to the background spectra in the CTH experiment than in DIII-D. Measured tungsten spectral lines in CTH are stronger, due to the greater solid angle of W emission as viewed by the CTH collection optics. Yet, the larger W signal to noise ratio in the CTH experiment significantly improves the confidence of W line identification in the DIII-D spectrum. The high density of W emission lines in the UV region allows for the possibility of simultaneous S/XB measurements and evaluating the effect of metastable levels on the relative intensities. Additionally, electron temperature and density measurements at the plasma boundary interface can also be determined by taking ratios of W I lines.

Numerous W I lines are identified between the CTH and DIII-D experiments in the range of 250 to 275 nm with a selection highlighted in Figure 3. Only three of the lines in this range (255.13 nm, 266.28 nm, and 268.14 nm) have been previously observed experimentally. Additionally, we present the first observation and identification of a particularly intense W I line at 265.65 nm that does not appear to be blended with other background emission. Emission from the CTH experiment in Figure 3 was acquired with the W-tipped probe fully inserted into the plasma while the DIII-D spectrum results from observation of a W coated insert on the divertor shelf. The same W I lines observed in both devices increases the confidence that the identified lines are primarily W and not blended with other impurities. The multiple strong W I lines existing in this region provides ample opportunity for erosion diagnosis. However, the corresponding tungsten erosion rate is not determined from this spectra because an absolute intensity calibration of the spectrometer for this wavelength range was not possible. Electron temperatures in the DIII-D divertor region between Edge Localized Modes (ELMs) mainly varied from 25-50 eV, with densities between $1 - 3 \times 10^{19}$ m$^{-3}$. Electron temperatures at the probe tip in CTH plasmas are not well known and certainly varied with probe insertion depth but are expected to be less than 50 eV. Electron densities around the probe in CTH ranged from 3 to 7 $\times 10^{18}$ m$^{-3}$ for the data shown. Another cluster of W I lines that has not been previously reported in fusion relevant plasmas, exists in the region between 310 and 340 nm. The spectra in Figure 4 was acquired during plasma conditions similar to Figure 3. The six lines identified in Figure 4 have not been previously observed in any fusion relevant plasma experiments. Only one of the lines could be clearly distinguished in DIII-D (321.6 nm) with the UV survey spectrometer. Nevertheless, the wavelength range in Figure 4 could be very promising for future high wavelength resolution measurements. Moreover, the longer UV wavelengths have the benefit of increased transmission in optical fibers as well as higher quantum efficiency for widely available CCD detectors.

A summary of all the W I spectral lines achieving an 'A', 'B', or 'C' rating between

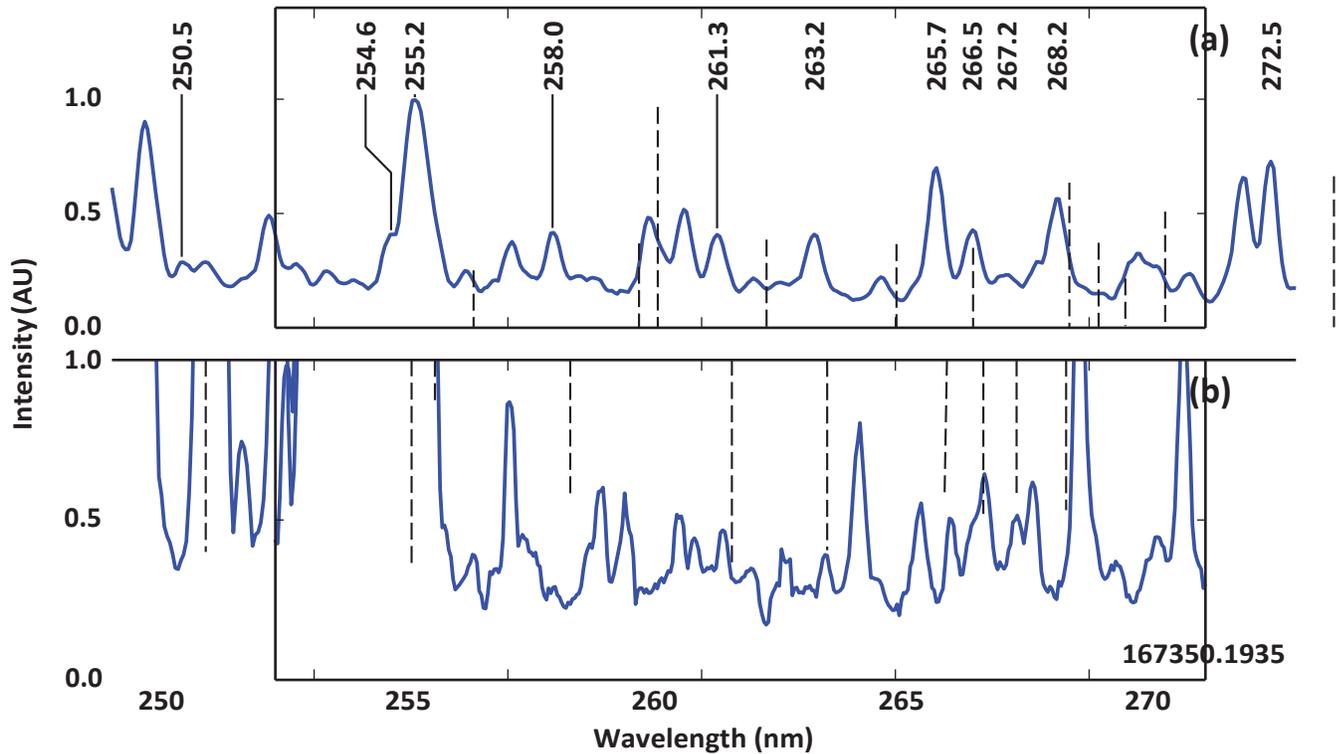

**Figure 3.** Measured CTH and DIII-D spectra with the wavelengths for neutral tungsten lines identified. (a) Average CTH spectrum of three shots (16120944, 16120947, 16120952) from 30 ms exposures beginning 60 ms into each discharge with the W probe fully inserted into the plasma. (b) DIII-D spectrum viewing the divertor shelf W coated inserts resulting from a single 200 ms exposure starting at 1935 ms into shot 167350.

the two experiments is provided in Table 1 along with full level identifications where possible. The level identification for the upper excited states is limited to $J$ values [29]. Table 1 includes first possible level identifications based on a structure calculation [23] for many of the excited states that produce strong spectral emission lines. Previous level identifications have been proposed [24]. Upper level identifications are obtained by combining the NIST Atomic Spectra Database with a prior structure calculation [23] but in some cases the upper level identification can not be uniquely determined so a partial identification is provided. For these cases, empty square brackets are used with the $J$-value as a subscript ($[\ ]_J$). The two observed emission lines at 233.23 and 294.69 nm are most likely multiplets which can not be resolved by the UV survey spectrometers. Therefore, the individual components of each multiplet are identified separately but the exact composition of the observed multiplet is not known. The table also indicates which lines have been previously observed in other fusion relevant experiments. It should be noted that the lowest six energy levels of neutral tungsten are long-lived states (either ground or metastable states) and could have quite large populations. These are the 5 $J$-values of the $5d^46s^2$ ($^5D$) term and $5d^56s$ ($^7S_3$) level. It is interesting that the majority of the observed lines in Table 1 have a lower level that is one of these six lower states

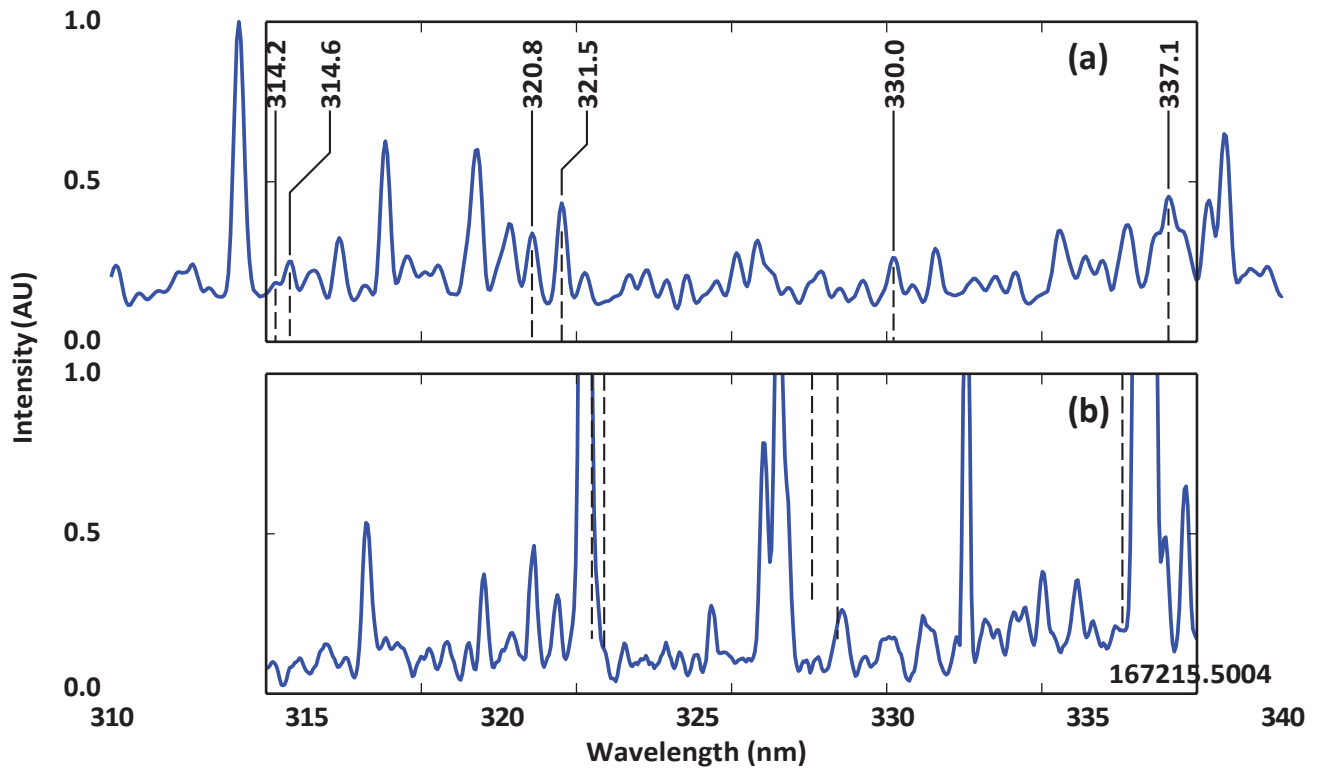

**Figure 4.** Measured CTH and DIII-D spectra with the wavelengths for neutral tungsten lines identified. (a) Average CTH spectrum of three shots (16120863, 16120864, 16120865) from 100 ms exposures integrating over each entire discharge with the W probe fully inserted into the plasma. (b) DIII-D spectrum viewing the divertor shelf W coated inserts resulting from a single 50 ms exposure starting at 5000 ms into shot 167215.

suggesting the observed lines are being mostly populated directly from these levels.

Probe experiments in CTH allow high-Z tipped probes to be inserted into the plasma at varying depths. CTH also employs a Radial Field Coil (RFC) which can be used to push the plasma down and away from the fully retracted probe, essentially moving the plasma boundary even further from the probe tip. Figure 5 highlights a selection of results from W-tipped probe experiments in CTH with varying amounts of plasma-probe interaction by adjusting the distance of the probe tip from the CTH midplane and the RFC magnitude. In general, as the probe is inserted into the plasma towards the midplane, the W lines become more intense and is most notable for the 250.47, 265.65, and 400.88 nm lines. The increase in W emission is consistent with more sputtering from the W-tipped probe closer to the plasma core. Other large impurity lines such as C IV and C III also present in Figure 5 demonstrate a similar trend with probe insertion but the impurity intensities do not change as rapidly as the W lines. Furthermore, when the probe is fully retracted (25.9 cm from the CTH midplane) and the RFC maximally energized (RFC high), the W line emission completely disappears while other impurity lines remain at a reduced intensity. The increase in intensity of lines other than W is attributed to greater plasma-probe interaction with higher density

**Table 1.** Table of W I emission lines observed between the DIII-D and CTH experiments along with level identifications where possible and confidence. Previously observed lines are indicated by superscripts on the NIST defined wavelengths: [a] [8], [b] [24], [c] [9], [d] [19], [e] [30], and [f] [31].

| Observed Wavelength [nm] | NIST Wavelength [nm] | Upper Level | Lower Level | Confidence |
|---|---|---|---|---|
| 233.23 | 233.28 | $5d^46s6p$ $(^3I_5)$ | $5d^46s^2$ $(^5D_4)$ | B |
|  | 233.28 | $5d^46s6p$ $(^5P_3)$ | $5d^46s^2$ $(^5D_3)$ | B |
| 243.48 | 243.60 | $5d^46s6p$ $(^5D_4)$ | $5d^46s^2$ $(^5D_3)$ | B |
| 246.24 | 246.28 | $([\ ]_1)$ | $5d^46s^2$ $(^5D_1)$ | C |
| 247.25 | 247.41 | $5d^46s6p$ $(^5F_4)$ | $5d^46s^2$ $(^5D_4)$ | B |
| 250.45 | 250.47 | $5d^46s6p$ $([\ ]_2)$ | $5d^46s^2$ $(^5D_1)$ | B |
| 254.61 | 254.71 | $5d^46s6p$ $(^5F_1)$ | $5d^46s^2$ $(^5D_2)$ | B |
| 255.15 | 255.13[a,b,d] | $5d^46s6p$ $(^3P_1)$ | $5d^46s^2$ $(^5D_0)$ | A |
| 256.18 | 256.20 | $([\ ]_3)$ | $5d^46s^2$ $(^5D_3)$ | C |
| 257.95 | 258.05 | $5d^46s6p$ $(^5P_1)$ | $5d^46s^2$ $(^5D_1)$ | A |
| 261.28 | 261.31 | $5d^46s6p$ $(^5F_2)$ | $5d^46s^2$ $(^5D_2)$ | B |
| 263.24 | 263.31 | $([\ ]_1)$ | $5d^46s^2$ $(^5D_1)$ | C |
| 265.69 | 265.65 | $5d^56p$ $(^7P_4)$ | $5d^5(^6S)6s$ $(^7S_3)$ | A |
| 266.46 | 266.28[b] | $([\ ]_2)$ | $5d^46s^2$ $(^5D_2)$ | C |
| 267.15 | 267.15 | $5d^46s6p$ $(^3F_3)$ | $5d^46s^2$ $(^5D_3)$ | B |
| 268.18 | 268.14[a,d] | $5d^46s6p$ $(^5G_4)$ | $5d^5(^6S)6s$ $(^7S_3)$ | A |
| 272.51 | 272.44 | $5d^56p$ $(^7P_3)$ | $5d^5(^6S)6s$ $(^7S_3)$ | A |
| 284.89 | 284.80 | $5d^46s6p$ $(^3D_3)$ | $5d^5(^6S)6s$ $(^7S_3)$ | B |
| 294.69 | 294.44[b,c] | $5d^56p$ $(^7P_2)$ | $5d^5(^6S)6s$ $(^7S_3)$ | B |
|  | 294.70[b,c] | $5d^46s6p$ $(^5F_3)$ | $5d^5(^6S)6s$ $(^7S_3)$ | B |
|  | 294.74[b,c] | $5d^46s6p$ $(^5I_4)$ | $5d^46s^2$ $(^5D_3)$ | B |
| 301.74 | 301.74 | $5d^46s6p$ $(^5D_4)$ | $5d^5(^6S)6s$ $(^7S_3)$ | B |
| 314.20 | 314.14 | $5d^46s6p$ $(^5F_4)$ | $5d^46s^2$ $(^3H_4)$ | B |
| 314.57 | 314.52 | $5d^46s6p$ $(^3H_5)$ | $5d^46s^2$ $(^3H_5)$ | B |
| 320.79 | 320.83 | $5d^46s6p$ $(^5F_2)$ | $5d^46s^2$ $(^5D_2)$ | B |
| 321.53 | 321.56 | $5d^46s6p$ $(^5F_5)$ | $5d^46s^2$ $(^5D_4)$ | A |
| 330.03 | 330.08 | $5d^46s6p$ $(^5F_4)$ | $5d^46s^2$ $(^5D_3)$ | B |
| 337.09 | 337.10 | $5d^46s6p$ $(^5F_2)$ | $5d^46s^2$ $(^5D_3)$ | B |
| 361.80 | 361.75[b,e] | $5d^46s(^6D)6p$ $(^5P_3)$ | $5d^5(^6S)6s$ $(^7S_3)$ | A |
| 378.08 | 378.08[b] | $5d^46s(^6D)6p$ $(^5P_2)$ | $5d^5(^6S)6s$ $(^7S_3)$ | B |
| 383.54 | 383.51[b] | $5d^46s(^6D)6p$ $(^5P_2)$ | $5d^46s^2$ $(^5D_2)$ | B |
| 386.81 | 386.80[b] | $5d^46s(^6D)6p$ $(^7D_4)$ | $5d^5(^6S)6s$ $(^7S_3)$ | B |
| 400.96 | 400.88[a,d,f] | $5d^46s6p$ $(^7P_4)$ | $5d^5(^6S)6s$ $(^7S_3)$ | A |

‡Only the upper level configuration and $J$-value are identified for the 250.47 nm line, while only the upper level $J$-value is identified for the 246.28, 256.20, 263.31, and 266.28 nm lines.

and temperature plasma parameters toward the core.

The tungsten probe scan experiment in CTH was utilized to confirm all W lines listed in Table 1. Identical probe scans were completed in CTH using both Mo and stainless steel probe tips. Quite notably, no lines were observed at 400.88, 265.65 or 255.13 nm during the Mo and stainless steel probe scans. Yet, when the W probe is fully inserted these are some of the strongest spectral lines. The background emission dependence on probe depth was investigated using Mo and stainless steel probe tips as there is no source of W in the CTH device. Comparison of the emission from the three probe tips allowed for persistent impurity lines to be distinguished from the W emission lines.

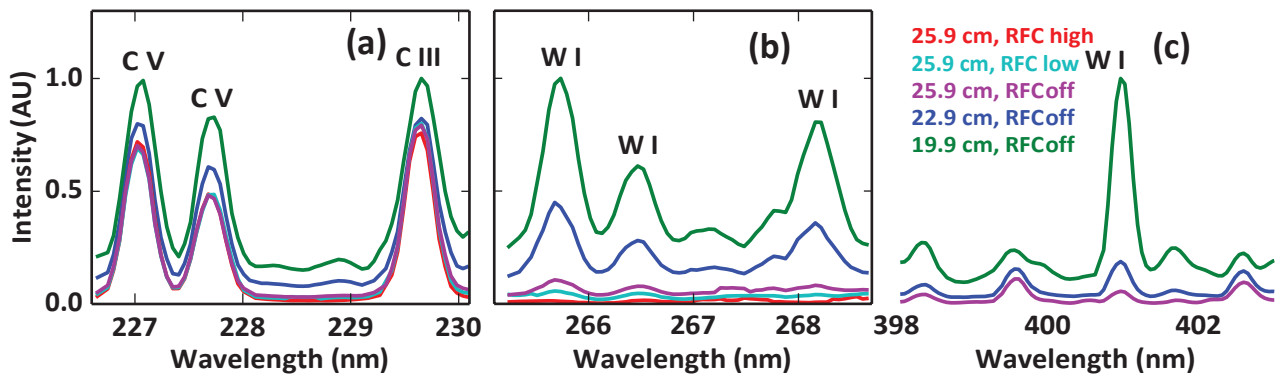

**Figure 5.** Intensities of tungsten and carbon impurity lines with varying W-tipped probe depth and Radial Field Coil (RFC) magnitude. The combinations of probe tip distance from the CTH midplane and RFC magnitude are listed in the legend from least plasma-probe interaction (25.9 cm, RFC high) to most interaction (19.9 cm, RFC off). (a) Carbon impurity line intensities increase modestly with probe insertion but are present for all configurations. (b) and (c) Tungsten lines demonstrate a significant dependence on probe depth and RFC magnitude. All tungsten lines are not present when the W-tipped probe is fully retracted with maximum applied radial field (25.9 cm, RFC high).

Identification of W line emission was achieved in the DIII-D experiment by observing transitions of the divertor strike point location from carbon to W coated inserts. Figure 6 depicts the time evolution of DIII-D shot 167350 with the UV spectrometers viewing a W insert on the divertor shelf for exposure times of 200 ms. During this discharge the outer strike point moves from the outboard portion of the lower divertor to the W coated inserts on the shelf where it is held fixed for approximately 3 seconds. Electron temperature and density on the W surface are measured by a single Langmuir probe embedded in the divertor shelf tile at the same radial location as the UV spectrometer collection spot [32]. The raw Langmuir probe data is plotted along with a trace having the ELM periods removed and smoothed over 101 points. Intensities of the W I, C II, and O V emission lines observed by the UV survey spectrometer throughout the discharge are binned using a 0.2 nm window encompassing the full width at half maximum of each spectral line and are shown normalized to their maximum intensity.

The intensities of all lines increase when the strike point is swept over the W insert; yet, the W I line increases later than the C II and O V impurity lines. Additionally, the ratio of the W I to C II emission also doubles when the strike point is swept over the W insert, where a ratio of 0.1 corresponds to the noise floor of the W I line at 265.65 nm while a ratio of 0.2 corresponds to measurable W I signal. The increase in W I lines relative to C lines assisted in discriminating W I radiation from other impurity lines.

A correlation analysis of W line intensities for UV wavelengths in DIII-D with various plasma parameters including ELM intensity has been attempted, as was previously reported in JET and DIII-D using the 400.88 nm line [28, 30]. However, the exposure time of the UV survey spectrometers (typically 100 ms) was such that multiple ELMs occurred during a single exposure. Therefore, signals from a $D_a$ filterscope system [33] were used as a proxy for the ELM intensity. The correlation analysis found no conclusive trend between multiple W I line intensities (including the 400.88 nm line) as measured by the UV survey spectrometers and $D_a$ emission. The correlation analysis was repeated using a high resolution visible spectrometer [34] viewing only the 400.88 nm and a correlation with the $D_a$ intensity was confirmed that is similar to the correlation found in [35]. The limited spectral resolution and throughput of the UV survey spectrometers likely resulted in lower signal to noise ratios that masked any correlated dynamics in the W I emission. It is anticipated that a higher-resolution spectrometer, sensitive to UV wavelengths, would similarly detect a correlation between the W I lines identified in Table 1 and ELM strength.

## 4. Conclusions

Ultraviolet survey spectrometers have been utilized to investigate emission from tungsten coated inserts in the DIII-D divertor and a tungsten-tipped probe in the CTH experiment for the potential benefit of erosion diagnosis. The radiated power of neutral tungsten is predicted to be most intense in the UV region and a total of 29 W I emission lines are observed and identified between the two experiments. Of the observed W I lines, 20 have not previously been reported in fusion relevant plasmas. The set of newly identified W I lines includes an intense line at 265.65 nm which is driven from the same metastable level as the widely utilized 400.88 nm. Thus, the 265.65 nm emission line could be important for benchmarking gross erosion measurements resulting from the 400.88 nm line. Material erosion can be determined from spectral line intensities together with atomic coefficients representing the 'ionizations per photon' (S/XB). However, the W I spectrum and any erosion measurement utilizing the S/XB technique is likely affected by the metastable level populations. Yet, simultaneous observations of multiple lines in combination with accurate atomic data could determine the ratios of the ground and all metastable populations.

For the UV lines confirmed in Table 1, all of the transitions decay to a ground or metastable state where the final level is one of the six lowest levels. Thus, it may be that for the strong UV lines observed, the initial level is populated primarily from a single driving metastable state. If that is the case, the spectral lines identified in Table 1 could thus prove to be very useful in determining ground and metastable fractions, their impact on the spectrum, and ultimately erosion measurements. In the region between 250 and 275 nm (Figure 3) there exist lines that transition to each of the five W I metastable levels. Moreover, the line at 255.13 nm is the only observed strong line that transitions to the ground level and thus is needed to carry out a complete metastable analysis. In the region between 310 and 340 nm (Figure 4) there exist lines that transition to the five W I metastable states but no measurable line that transitions to the ground state. Future work is needed to quantify the impact of metastable fraction on the observed W I spectrum for different plasma conditions and the resulting erosion diagnostics. The lines presented in this work provide a mechanism for studying the role of metastable populations on S/XB diagnostics and have the potential to produce more accurate tungsten total erosion measurements.

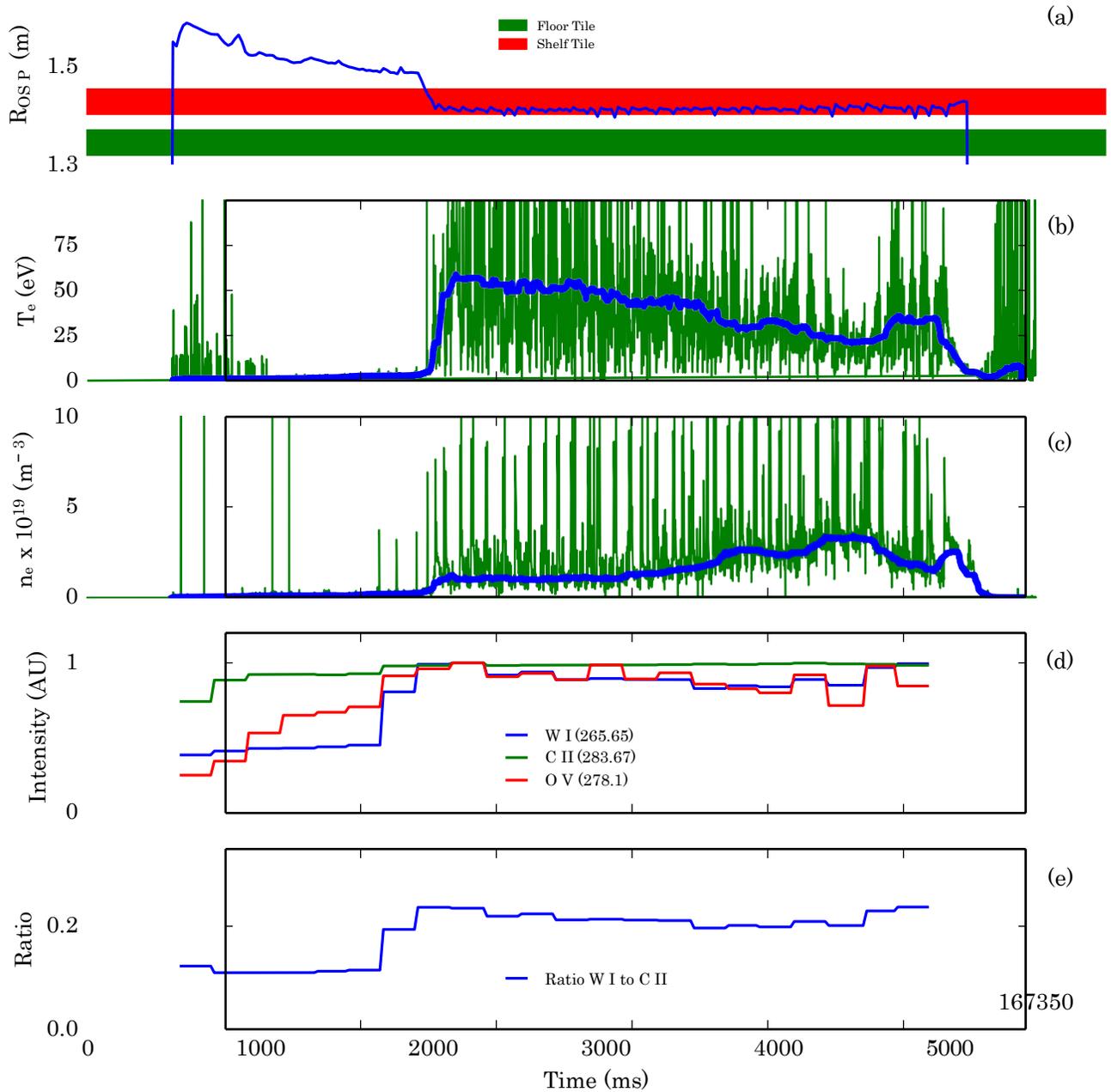

**Figure 6.** Time evolution of DIII-D shot 167350: (a) Radial position of the outer strike point ($R_{OSP}$) with the locations of the floor and shelf W inserts shown in green and red respectively, (b) and (c) raw (green) and ELM-free smoothed (blue) electron temperature and density on the W surface measured by a Langmuir probe, (d) W I, C II, and O V line intensities normalized to their maximum intensity during the discharge, (e) ratio of W I to C II line intensities.

## 5. Acknowledgements

This work is supported by the U.S. Department of Energy, Office of Science, Office of Fusion Energy Sciences, using the DIII-D National Fusion Facility, a DOE Office of Science user facility, under Awards DE-SC0015877, DE-FC02-04ER54698, DE-FG02-00ER54610, and DE-AC52-07NA27344. DIII-D data shown in this paper can be obtained in digital format by following the links at https://fusion.gat.com/global/D3D DMP.

## 6. Disclaimer